\documentclass[
    pre,
    twocolumn,
    twoside,
    showkeys,
    byrevtex,
    superscriptaddress,
    floatfix,
    nofootinbib,
    longbibliography
]{revtex4-1}

\usepackage{settings}

\setboolean{twocolswitch}{true}
\begin{document}

\title{\protect 
Quantifying language changes surrounding mental health on Twitter

}
\author{
\firstname{Anne Marie}
\surname{Stupinski}
}
\email{astupins@uvm.edu}
\affiliation{
  Computational Story Lab,
  Vermont Complex Systems Center,
  University of Vermont,
  Burlington, VT 05405.
} 

\author{
\firstname{Thayer}
\surname{Alshaabi}
}
\affiliation{
  Computational Story Lab,
  Vermont Complex Systems Center,
  University of Vermont,
  Burlington, VT 05405.
} 

\author{
  \firstname{Michael V.}
  \surname{Arnold}
}
\affiliation{
  Computational Story Lab,
  Vermont Complex Systems Center,
  University of Vermont,
  Burlington, VT 05405.
}

\author{
  \firstname{Jane Lydia}
  \surname{Adams}
}
\affiliation{
  Computational Story Lab,
  Vermont Complex Systems Center,
  University of Vermont,
  Burlington, VT 05405.
} 

\author{
  \firstname{Joshua R.}
  \surname{Minot}
}
\affiliation{
  Computational Story Lab,
  Vermont Complex Systems Center,
  University of Vermont,
  Burlington, VT 05405.
}

\author{
  \firstname{Matthew}
  \surname{Price}
}
\affiliation{
  Department of Psychology,
  University of Vermont,
  Burlington, VT 05405.
}

\author{
  \firstname{Peter Sheridan}
  \surname{Dodds}
}
\affiliation{
  Computational Story Lab,
  Vermont Complex Systems Center,
  University of Vermont,
  Burlington, VT 05405.
} 
\affiliation{
  Department of Computer Science,
  The University of Vermont,
  Burlington, VT 05405.
}
\affiliation{
  Department of Mathematics \& Statistics,
  The University of Vermont,
  Burlington, VT 05405.
}

\author{
  \firstname{Christopher M.}
  \surname{Danforth}
}
\email{cdanfort@uvm.edu}
\affiliation{
  Computational Story Lab,
  Vermont Complex Systems Center,
  University of Vermont,
  Burlington, VT 05405.
} 
\affiliation{
  Department of Mathematics \& Statistics,
  The University of Vermont,
  Burlington, VT 05405.
}
\affiliation{
  Department of Computer Science,
  The University of Vermont,
  Burlington, VT 05405.
}

\date{\today}

\begin{abstract}
  \protect
  Mental health challenges are thought to afflict around 10\% of the global population each year, with many going untreated due to stigma and limited access to services. 
Here, we explore trends in words and phrases related to mental health through a collection of 1- , 2-, and 3-grams parsed from a data stream of roughly 10\% of all English tweets since 2012. 
We examine temporal dynamics of mental health language, finding that the popularity of the phrase `mental health’ increased by nearly two orders of magnitude between 2012 and 2018. 
We observe that mentions of `mental health' spike annually and reliably due to mental health awareness campaigns, as well as unpredictably in response to mass shootings, celebrities dying by suicide, 
and popular fictional stories portraying suicide.
We find that the level of positivity of messages containing `mental health', while stable through the growth period, has declined recently.
Finally, we use the ratio of original tweets to retweets to quantify the fraction of appearances of mental health language due to social amplification. 
Since 2015, mentions of mental health have become increasingly due to retweets, suggesting that stigma associated with discussion of mental health on Twitter has diminished with time.

\end{abstract}



\maketitle


\section{Introduction}
\label{sec:introduction} 

Recent estimates place 1 in 10 people globally as experiencing from some form of mental illness~\cite{owidmentalhealth}, with 1 in 30 suffering from depression~\cite{who2020depression}. 
These rates put mental illness 
among the leading causes of ill-health and disability worldwide. 
Moreover, rates of mental health disorders and deaths by suicide have increased in recent years, especially among young people~\cite{mcclure2001suicide}. 

Since the beginning of the COVID-19 pandemic and the subsequent social isolation, there have been recordings of drastic declines in physical activity and time spent socializing, and coinciding increases in screen time and symptoms of depression~\cite{giuntella2021lifestyle}.
Google searches for mental health related topics also increased in the first weeks of the pandemic, leveling out after more information regarding stay-at-home orders were released~\cite{jacobson2020flattening}. 
Following March 2020, there has also been a measured 
increase in suicidal ideation 
that is associated with elevated reports of isolation~\cite{fortgang2021increase}.
The service Crisis Text Line reported receiving a higher than average volume of messages for every day following March 16th in the year 2020, with the main topics being anxiety, depression, grief, and eating disorders~\cite{crisistextline_2021}.
Price~\etal~\cite{price2021doomscrolling} also found that daily ``doomscrolling''---repeatedly consuming negative news and media content online---was associated with same-day increases in depression and PTSD. These effects were larger among those with a prior history of psychopathology and trauma exposure. 
The pandemic also influenced what content people discussed on social media, with users shifting away from ``self-focused'' perspectives and towards more ``other-focused'' topics that used to be vulnerable or taboo to discuss~\cite{nabity2020inside}.
A survey of American adults during the pandemic found that the depth of distressing self-disclosures posted online could be predicted by a user's perceived anonymity, visibility control, and closeness to their audience~\cite{zhang2021distress}.

Historically, the availability of mental health treatment services has not meet the demand for such~\cite{Detels}.
Mental health care also experiences a paradox of being over-diagnosed yet under-supported, with some symptoms and disorders being readily medicated despite not being understood and accepted socially~\cite{plos2013paradox}. 
Furthermore, many who would benefit from mental health services do not seek or participate in care, as they are either unaware of such services, are unable to afford them, or the stigma associated with seeking treatment proves too great a barrier~\cite{Corrigan}. 
In fact, two-thirds of people with a known mental disorder do not seek help from a health professional~\cite{world2001world}. 

While stigma has proven to be a significant barrier to receiving treatment from formal (e.g., psychiatrists, counselors) and informal sources (e.g., family and friends), the COVID-19 pandemic and subsequent isolation have spurred awareness of mental illness and discussion on this topic in public forums such as social media. Measuring changes in this conversation, we aim to quantify the hypothesized increase in discussions and awareness, and the corresponding reduction in stigma around mental illness. 
Our findings suggest that the number of mental health conversations on Twitter have substantially increased in recent years, particularly on dates associated with either awareness campaigns or tragedies. 
We also examine social attention and expressed happiness in an attempt to piece together how this conversation has shifted in the past decade.

Many researchers have used social media platforms in order to explore and understand dynamics of healthcare discussion~\cite{gohil2018sentiment}. 
Several reviews have been done on mental health discussion in particular, finding that social media is a viable platform for users to discuss mental health and feel supported, although privacy risks and ethical concerns of research applications exist as well~\cite{conway2016social,naslund2020social}. 
A study by De Choudhury \etal~\cite{de2013predicting} analyzed the Twitter activity of individuals diagnosed with depression, along with clinically validated measures, in order to predict users who may be at risk of the mental illness. 
Reece and Danforth~\cite{reece2017forecasting} used tweets posted prior to a user's diagnosis date to identify social media content associated with the onset of depression. 

De Choudhury \etal~has also worked on predicting postpartum depression in new mothers, using Facebook activity, linguistic expression in status updates, and demographic survey data~\cite{de2014characterizing}.
Using consenting Instagram users' photos, Reece \etal~\cite{reece2017instagram} identified distinct predictive markers of depression in the images posted by individuals previously diagnosed with depression by a psychiatrist.
Work by Coppersmith \etal~\cite{coppersmith2014quantifying} classified online users who suffer from Post Traumatic Stress Disorder by using self-disclosing messages on Twitter. 

Another study using self-disclosures proposes a classifier to distinguish between Twitter users suffering from mental illness from those who are not, using  messages collected from individuals self-reporting ten various mental illnesses~\cite{coppersmith2015adhd}. 
Using the self-disclosure classification methods proposed by Coppersmith \etal~\cite{coppersmith2014quantifying}, Bathina \etal~\cite{bathina2021individuals} reported that Twitter users with a diagnosis of depression show higher levels of distorted thinking in their posts when compared to a random sample of messages. 

Researchers have used mental health support threads on Reddit to examine the shift of suicidal ideation on social media, identifying users who are more likely than others to make this transition from the typical content in these threads~\cite{de2016discovering}. 
Analysis of text-based crisis counseling conversations found actionable strategies associated with more effective counseling, such as adaptability, dealing with ambiguity, creativity, and change in perspective~\cite{althoff2016large}.
While developments in predicting mental health states provide an opportunity for early detection and treatment, they come with several ethical concerns, such as incorrect predictions, involvement of bad actors, and potential biases~\cite{chancellor2019taxonomy}.
Social media users also hold negative attitudes towards the concept of automated well-being interventions prompted by emotion recognition, stating that any automated message could not have the personal, human attributes necessary for such an interaction to be successful~\cite{Roemmich2021Conceptualizations}. 
They also view emotion recognition in general as invasive, scary, and a loss of their control and autonomy, as people view emotions as insights to behavior that are vulnerable and prone to manipulation~\cite{andalibi2020human}.

Several other studies have more directly examined attitudes towards those with mental illnesses, attempting to measure the stigma towards these individuals that exists in social communities. 
Rose \etal~\cite{Rose} sought to investigate the extent of stigma and treatment avoidance in 14-year-old students in relation to how they refer to people with mental illness, finding that the majority of phrases used fit into the theme ``popular derogatory terms''. 

Reavley and Pilkington took a qualitative approach to monitoring stigma on Twitter, collecting tweets over a 7-day period that contain the hashtags \#depression or \#schizophrenia and categorizing them~\cite{Reavley}.
These tweets were coded based on the attitude they indicated (stigmatizing, personal experience, supportive, neutral, or anti-stigma) and on their content (awareness promotion, research findings, resources, advertising, news media, or personal opinion). 
Their findings show that tweets related to depression mostly contain resources or advertisements for mental health services, while tweets on schizophrenia contain awareness promotion or research findings. 
The percentage of tweets showing stigmatizing attitudes was 5\%, and most of these showed inaccurate beliefs about schizophrenia being multiple personality disorder. 

Another recent study by Robinson \etal~\cite{robinson2019measuring} used Twitter to investigate attitudes towards a variety of mental and physical health conditions, finding that mental health conditions were more stigmatized and trivialized than physical ones, especially among mentions of schizophrenia and obsessive compulsive disorder.
Examinations of Chinese social media posts on the platform Weibo find that roughly six percent of posts include stigmatizing attitudes towards depression, reflecting beliefs that depression is ``a sign of personal weakness'' as well as ``not a real medical illness''~\cite{li2018detecting}. 

The goal of our present study is to contribute to this growing body of work, previously largely focused on individuals, by using a data-driven approach to examine the collective conversation over a full decade.
Using messages from Twitter, we analyze the conversation around mental health by examining the growth of public attention, the divergence of language from general messages and the associated happiness shifts, and the rise of ambient words or phrases. 

We structure our paper as follows.    
In Sec.~\ref{sec:methods}, we describe in brief the mental health data set 
using the Storywrangler instrument~\cite{alshaabi2020storywrangler} for Twitter, 
which provides day-scale $n$-gram time series data sets for $n=1, 2, 3$. 
In Sec.~\ref{sec:results}, we explore several aspects of the conversation related to mental health on Twitter, such as the growth of collective attention to the topic and the associated ambient happiness (Sec.~\ref{subsec:attention}), and narrative and social amplification trends, 
looking into the specific language and retweet ratios of this dataset compared to general Twitter (Sec.~\ref{subsec:narrative}). 
In our concluding remarks, 
we outline several limitations of our study and some potential future developments in this work.

\begin{table*}[!tp]
\centering
\begin{tabularx}{\linewidth}{L||C|C|C|C|C|C}
& \multicolumn{2}{c|}{2012-02-08} &  \multicolumn{2}{c|}{2014-01-28} & \multicolumn{2}{c}{2018-01-31}\\
\hline
& MH & General & MH & General & MH & General \\
\hline\hline
Unique 1-grams & $3.0 \times 10^3$  & $1.7 \times 10^7$ & $1.6 \times 10^3$ & $2.4 \times 10^7$ & $4.9 \times 10^4$ & $2.1 \times 10^7$ \\
Total 1-grams & $3.0 \times 10^4$ & $3.1 \times 10^8$ &  $2.3 \times 10^4$ & $4.9 \times 10^8$ & $4.4 \times 10^6$ & $5.4 \times 10^8$  \\
Total 1-grams (no retweets) & $9.3 \times 10^3$ & $2.2 \times 10^8$ & $1.5 \times 10^5$ & $2.9 \times 10^8$ & $2.6 \times 10^5$ & $1.6 \times 10^8$  \\
\end{tabularx}
\caption{ \textbf{Summary statistics of the mental-health $n$-gram dataset compared to the general Twitter $n$-gram dataset on three individual days.} Dates shown correspond to `Bell Let's Talk' Day, an annual fundraising and awareness campaign, which is also coincident with the annual peak in conversation regarding mental health. \textit{Unique 1-grams} enumerate the set of distinct words found in tweets on these dates, reflecting roughly 10\% of all tweets. 
The \textit{total 1-grams} row shows the sum of counts of each unique 1-gram, and \textit{total 1-grams (no retweets)} is the sum of the counts of 1-grams in tweets not including any messages that were retweeted. In 2012, roughly 1 in 10,000 messages referenced mental health. In 2018, the rate increased to roughly 1 in 100 messages. }
\label{tab:ngrams}
\end{table*}


\section{Data and Methods}
\label{sec:methods}

\begin{figure*}[!tp]
    \centering
    \includegraphics[width=\textwidth]{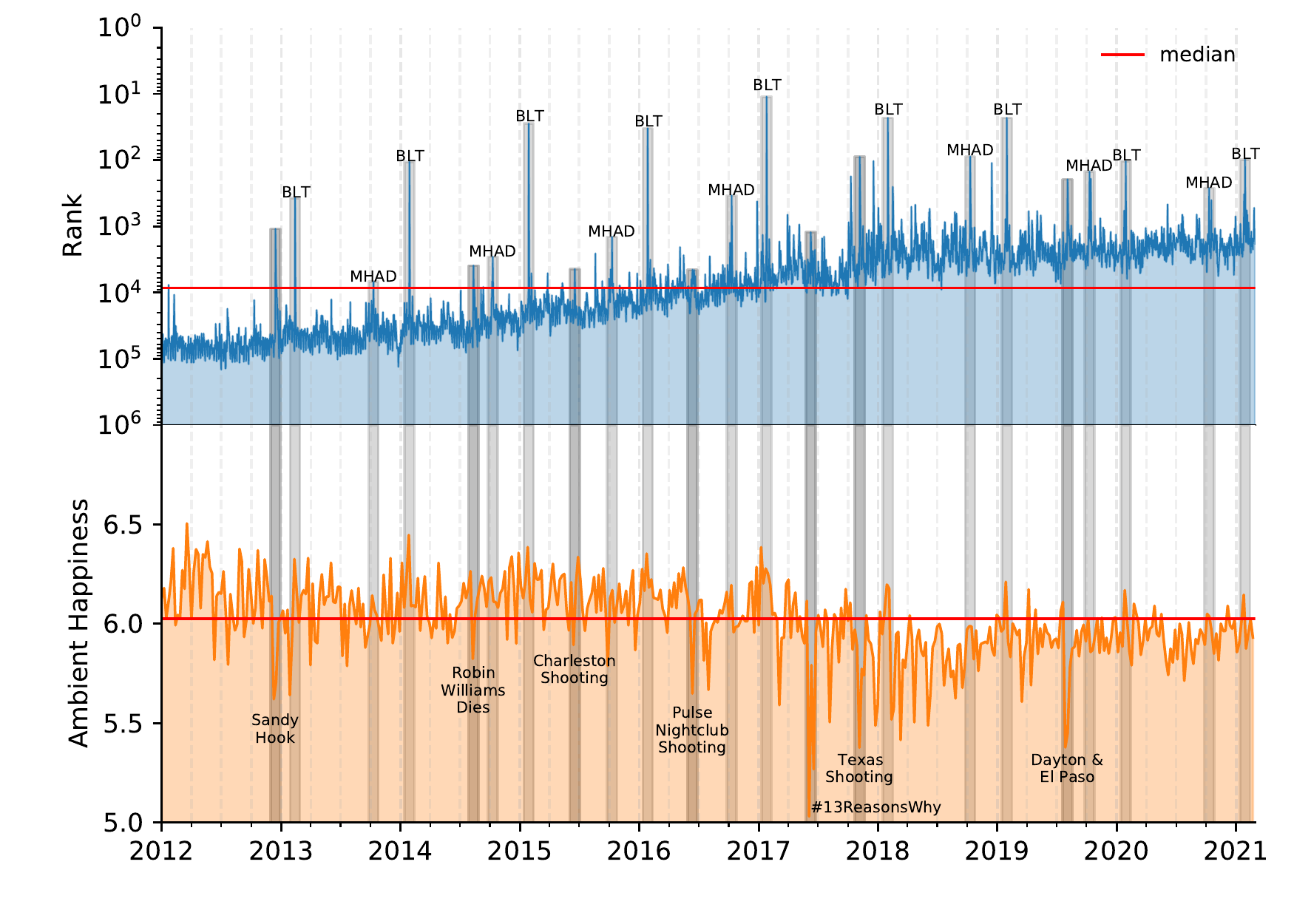} 
    \caption{\textbf{Timeline of mental health discourse on Twitter.} The top panel shows the rank timeseries of the 2-gram ``mental health'' over the past decade on a logarithmic axis. 
    Rank is determined by ordering 2-grams in descending order of counts for each day, and plotted on an inverted axis. 
    The median rank value of the timeseries is highlighted by a horizontal red line. 
    Between 2012 and 2018, the phrase increased in rank by nearly two orders of magnitude, reflecting a dramatic increase in the discussion of mental health on Twitter. 
    The bottom panel shows the ``ambient happiness'' of all messages containing the 2-gram ``mental health'' for each day over the same time period. 
    For clarity, this data is shown as a weekly rolling average, and again the median is highlighted by a red horizontal line. 
    Ambient happiness remained roughly constant during the period of increasing volume, but has dropped since 2017.
    Across both panels, key dates are highlighted in grey and annotated with the associated event. 
    These are dates that led to large spikes or drops in either timeseries. 
    Annually occurring events, such as Bell Let’s Talk (BLT) or Mental Health Awareness Day (MHAD), are shown with light grey, and unexpected events are highlighted with a darker grey. 
    Ambient happiness dips tend to correspond to mass shootings, with the lowest period coinciding with discussion of the Netflix series ``13 Reasons Why.''
}
    \label{fig:MHtimeseries}
\end{figure*}

\begin{figure*}[!tp]
    \centering
    \includegraphics[width=\textwidth]{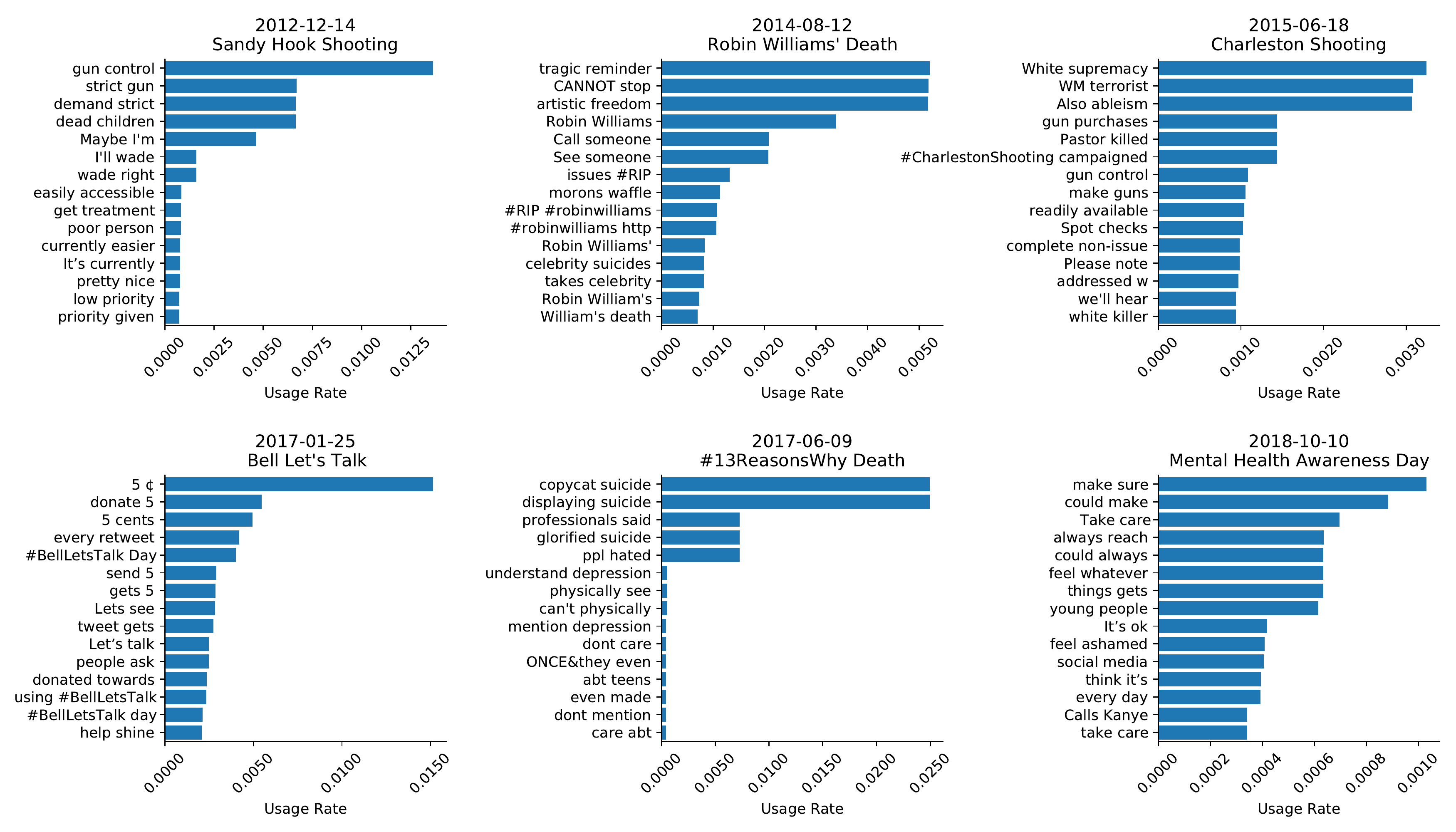} 
    \caption{\textbf{Top $n$-grams used in discussions of mental health.} Here we show the top fifteen 2-grams that appear in the ``mental health" tweet collection for a few outlier dates noted in Fig.~\ref{fig:MHtimeseries}. Each subplot lists the date and its associated event, along with a bar graph of the usage rate. It is worth noting that the bars in each subplots cannot be compared to those of the other subplots, as the range of the x-axes are varied for clarity.
    }
    \label{fig:ngrams_grid}
\end{figure*}

Twitter is a real-time source of information on a wide variety of topics. Since most tweets are public and the platform is commonly used by both adults and young people, estimates of public opinion based on the platform can complement survey-based measures.
Complicating this effort, Twitter's user base is limited~\cite{perrin2019share}, skewing slightly younger and more politically left-leaning than the US population overall. For these reasons and many others, Twitter messages will fail to capture many aspects of human behavior. 

In particular, mental health discourse is a sensitive, often personal topic that many individuals will avoid discussing publicly.
Nevertheless, Twitter is a valuable social ecosystem from which we can sketch a rough portrait of the existing conversation around mental health. And given that social media lowers the barrier for individuals to join difficult conversations, especially with Twitter allowing users to sign up anonymously, it is a promising source of unstructured language data describing the changing experience of a stigmatized group.

The source of data for the present study is Twitter's Decahose API, filtered for English messages, from which we collect a 10\% random sample of all public tweets between January 2012 and January 2021. This collection is separated into three corpora consisting of (a) all tweets, (b) tweets containing the phrase ``mental health'', and (c) tweets containing a small set of phrases related to mental health. Statistics and timeseries comparisons between corpora are made as follows.

To explore trends in the appearance of words, we process messages into 1-, 2- and 3-grams, where a 1-gram is a one-word phrase, 2-gram is a two-word phrase, and so on, using the $n$-gram popularity dataset Storywrangler~\cite{alshaabi2020storywrangler}. 

For each day, we count the number of times each unique $n$-gram appears in tweets, and determine usage frequencies relative to the appearance of other phrases on Twitter. 
We rank $n$-grams by descending order of count;
$n$-grams with a low rank value assigned to phrases appear on Twitter very often, while those with a high rank value appear rarely.
For example, the 1-gram `a' has a median rank of 1, as it is typically the most commonly used word in the English language.
Meanwhile, the 1-gram `America' is less common, with a median rank of 990 \cite{dodds2019fame}.
In order to better visualize this concept of descending count in the figures to follow, we will plot rank on an inverted axis.

To explore the specific language used when discussing mental health on Twitter, we compile a separate collection of $n$-grams from tweets related to this topic. 
Restricting to messages that contain the 2-gram ``mental health", we create $n$-grams in the same fashion as previously described, determining their usage frequency within this anchor set and ranking phrases by descending order of counts. 
Summary statistics for key dates in this new dataset compared to the general 1-grams dataset are shown in Table~\ref{tab:ngrams}. 
We also compute the aggregated frequency and rank of $n$-grams over each year.

Using these datasets, namely counts of phrases in all tweets (General) vs. counts of phrases in tweets containing ``mental health'' (MH), we analyze changes in the conversation surrounding mental health through time. The dynamics of several other phrases related to mental health are analyzed as well, but we focus primarly on ``mental health'' as a representative example of such phrases rather than attempting to exhaustively gather all related content.


\section{Results and Discussion}
\label{sec:results}

\begin{figure*}[!tp]
    \centering
    \includegraphics[width=.8\textwidth]{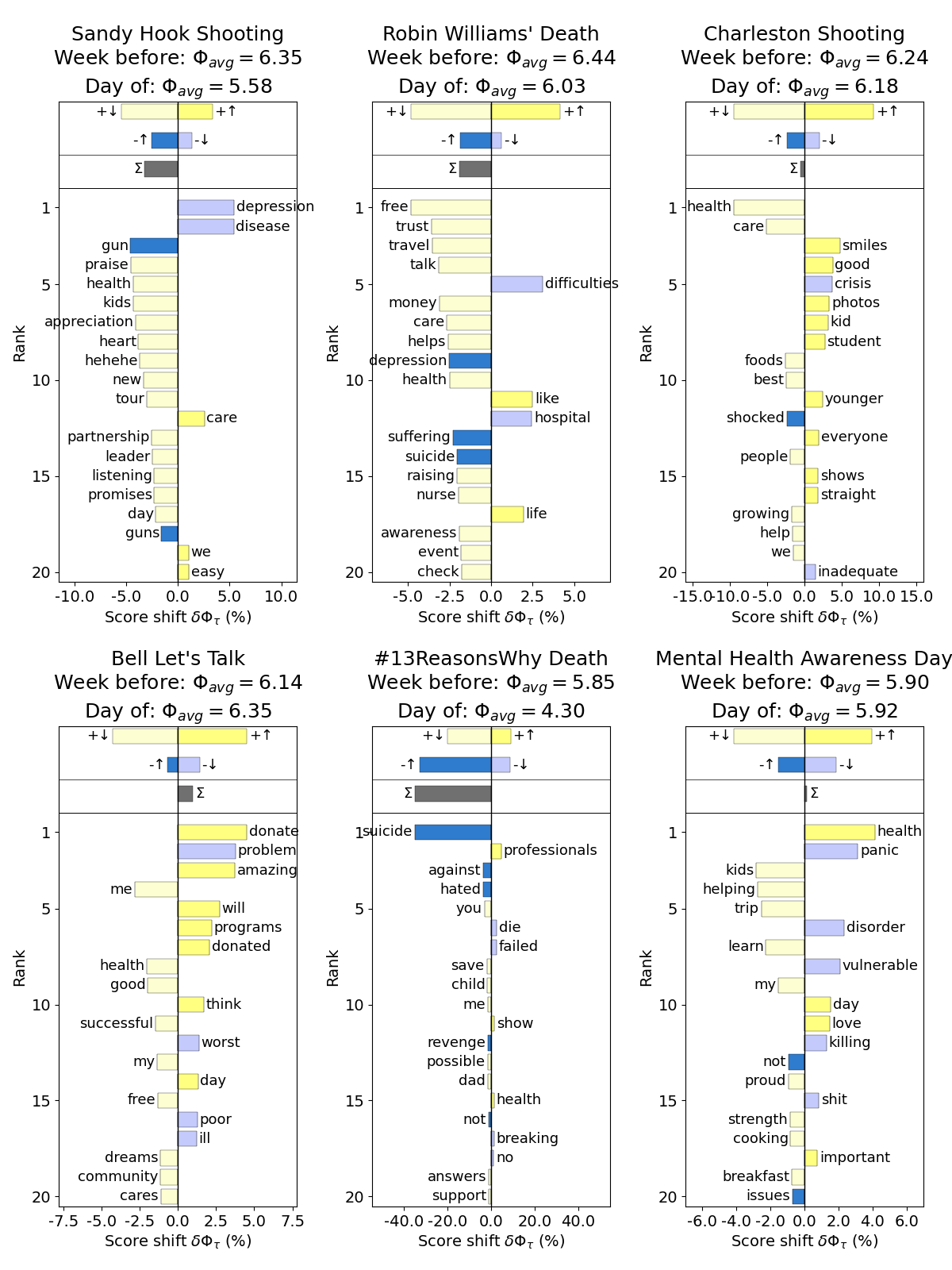} 
    \caption{
    \textbf{Happiness word shift graphs.}
    In each of the six panels, we show the twenty 1-grams that contribute most to the shift in ambient happiness on key dates shown in Fig.~\ref{fig:MHtimeseries}, relatively to the prior week. The words shown in blue are ones that have been labeled as relatively negative, and the ones shown in yellow have been labeled as relatively positive~\cite{gallagher2021generalized}. For example, on the day of the Sandy Hook shooting, the relatively negative word ``gun" appeared more often in mental health tweets than during the prior week, while the relatively negative words ``depression" and ``disease" appeared less often. On Bell Let's Talk day, the relatively positive words ``donate" and ``amazing" appear more often, and the relatively negative words ``problem" and ``worst" appear less often. The darker shade of these colors tells us where there is an increase in these words, while the lighter shade represents a decrease in usage. 
    The happiness score shift is shown on the horizontal axis, representing how positive or negative the language on these days becomes, and the happiness rank of the 1-gram in this subset is shown on the vertical axis. Average ambient happiness scores for the day of the event, as well as a week before the event, are also noted at the top of each subplot.
    }
    \label{fig:wordshift}
\end{figure*}

\subsection{Growth of Collective Attention}
\label{subsec:attention}

Public awareness and education on an issue is an important step in reducing negative attitudes, as a major component of stigma is a lack of knowledge~\cite{Corrigan}. 
In order to understand the general public’s level of awareness of mental health issues, we quantify the frequency at which people on Twitter have discussions about the topic of mental health. 
Using Twitter $n$-gram data, we construct a rank timeseries of the 2-gram ``mental health” on a logarithmic axis, which can be seen in Fig.~\ref{fig:MHtimeseries}. 

We find that this 2-gram increased in rank by nearly two orders of magnitude between 2012 and 2018. For the first four years, only a handful of dates resulted in ranks for ``mental health'' more popular than the overall median, while for the final four years, only a few dates result in ranks indicating less attention than the median.
This substantial increase is evidence that the conversation around mental health is happening far more frequently. 

We also examine the positivity of this conversation, calculating the ``ambient happiness" score of messages mentioning the phrase ``mental health” for each day, which is also shown in Fig.~\ref{fig:MHtimeseries}. 
Ambient happiness scores for each day are computed by averaging the scores of each word that appear in a message with ``mental health" for a given day, 
using the labMT dictionary~\cite{dodds2011temporal}. 
While the rank of this 2-gram has increased over the past decade, 
the ambient happiness of these messages have decreased.

Examining the daily behavior of these timeseries, several dates emerge where either the rank or ambient happiness deviate largely from their baseline behavior. 
In Fig.~\ref{fig:MHtimeseries}, key events associated with large spikes or drops in the timeseries are highlighted across both panels. 
Awareness events such as Bell Let’s Talk (BLT) and Mental Health Awareness Day (MHAD) contribute to the large, annual spikes in rank beginning in 2013. 

Bell Let’s Talk, falling on the last Wednesday of January each year, was started by the Canadian company Bell Telephones, and aims to bring awareness to the general public about mental health issues by donating five cents for each tweet using their hashtag `\#BellLetsTalk'. 
The 2-gram `mental health' reached its highest rank ever on Bell Let’s Talk day in 2017, peaking at the 18th most popular phrase compared to all other 2-grams on Twitter that day.

Other spikes in rank, and concurrent drops in ambient happiness, occurred on dates with national tragedies such as mass shooting events or celebrity deaths. 
The largest drop in ambient happiness occurred in 2017, immediately following the death of a teen that was connected to the Netflix series ``13 Reasons Why''~\cite{kindelan2california}.
Looking at the events that sparked more conversation around the topic of mental health, and their associated levels of ambient happiness, awareness campaigns tend to lead to a rise in ambient happiness, while the unexpected events, of which all would be considered tragedies, lead to drops in ambient happiness.

Looking further into the language used on these specific dates, we show the top $n$-grams found in messages containing ``mental health” in Fig.~\ref{fig:ngrams_grid}. 
These co-occurring $n$-grams are shown with their usage rate, rather than rank, so that we can visually see how phrases are being used compared to the others in the same list.

For example, a popular article shared on December 14, 2012 contained the phrase ``It’s currently easier for a poor person to get a gun than it is for them to get treatment for mental health issues." which was subsequently quoted by thousands of accounts on Twitter~\cite{mukherjee_2012}.
The resulting phrases seen in this figure provide more insight into what the broader conversation around mental health looks like following these events.

To understand the rise and fall of the ambient happiness scores over the timeseries in Fig.~\ref{fig:MHtimeseries}, we can look at the words that most heavily contribute to these shifts~\cite{gallagher2021generalized}. 
Fig.~\ref{fig:wordshift} highlights words associated with the same key events shown in Fig.~\ref{fig:ngrams_grid}, using messages from a week before the event as a reference. 
Words highlighted with a blue bar are ones that have been coded as negative, and words with a yellow bar have been coded as positive.

The darker shades of these two colors represent words that have increased in usage compared to the reference, while lighter shades represent words that have decreased in usage. 
The left side of these panels shows words that are bringing the average score down, whether with an increase in negative words or a decrease in positive words, and the right side shows words that are raising the score. 
The average ambient happiness scores for the day of the event and a week before the event are also highlighted at the top of each panel. 
The 1-grams are also ordered by rank from top to bottom, as shown by the vertical axis.

Looking at Fig.~\ref{fig:wordshift}, we see that mass shooting events have an increase in negative words such as ``gun'', ``guns'', and ``shocked'', and a diminishing use of negative words such as ``depression'', ``disease'', and ``crisis''. 
The day of the Sandy Hook shooting saw less positive words such as ``praise'', ``appreciation'', and ``listening'', which would usually be seen in the daily mental health content on Twitter.

While the Charleston shooting saw a decrease in words such as ``health'' and ``care'', it also saw an increase in positively coded words such as ``smiles'', ``kid'', and ``student'', which likely refer to the shooter in this event. 
This example highlights the drawbacks of dictionary-based 
ambient happiness analysis without context of the words being used, as independently positive words can be used to describe a tragic event and vice versa. 
The middle panels in both rows highlight word shifts following death by suicide tragedies, and include an increase in the words ``depression'', ``suffering'', and ``suicide'', 
which explain the drops in ambient happiness seen on these days.

The awareness events Bell Let's Talk and Mental Health Awareness Day, which represent the only increases in ambient happiness of the dates shown in Fig.~\ref{fig:wordshift}, both show an increase in quite a few positive words: ``donate'', ``amazing'', ``programs'', ``health'', ``love'', and ``important''. 
These days also notably see a decrease in strongly negative words, such as ``problem'', ``disorder'', ``vulnerable'', and ``killing''. 
These results highlight the shift in language on awareness days, away from phrases with negative connotations and focusing on language relating to community support and aid.


\begin{figure*}[!tp]
    \centering
    \includegraphics[width=\textwidth]{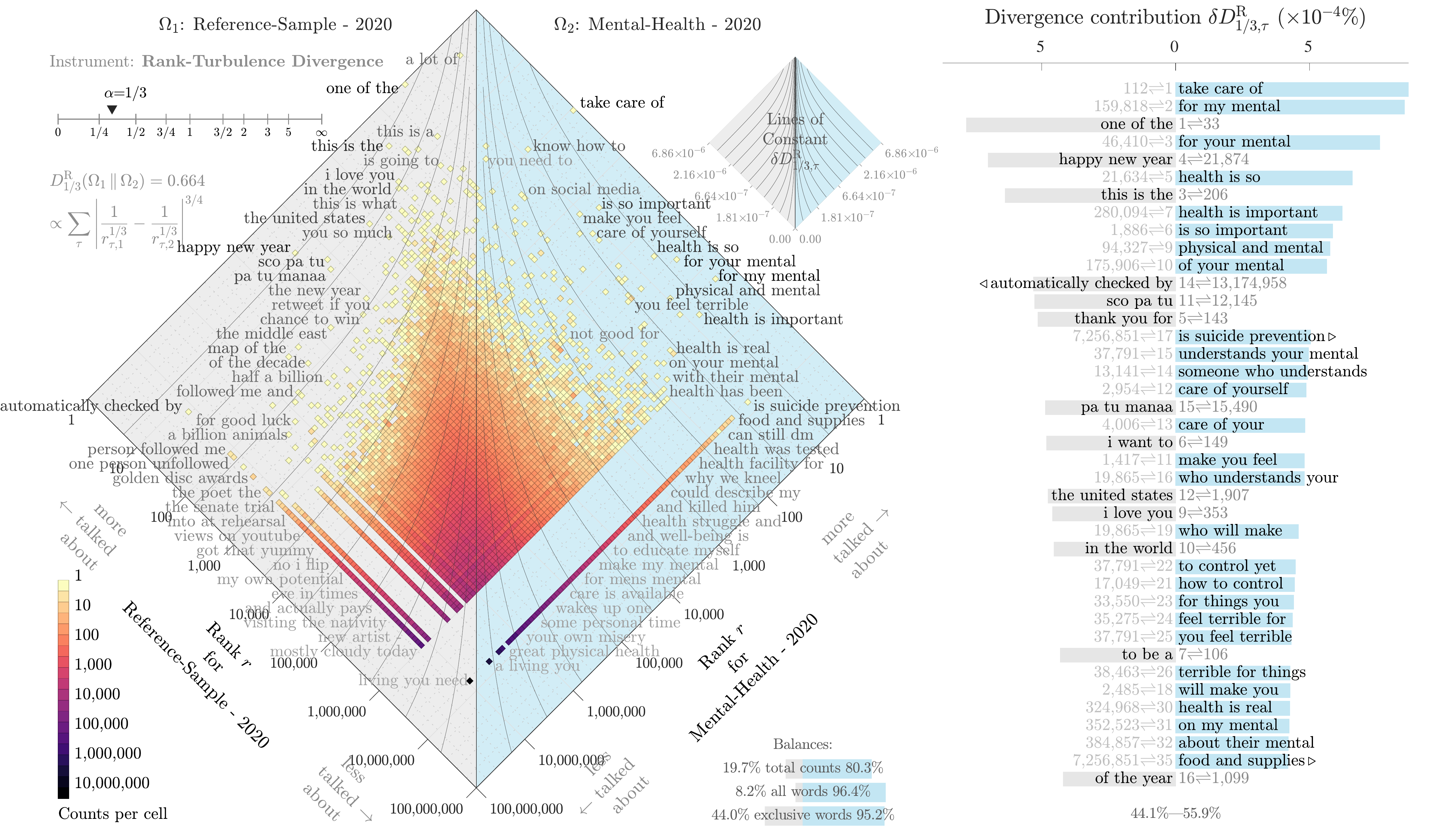} 
    \caption{
    \textbf{Allotaxonograph using rank-turbulence divergence of 1-grams from tweets in 2020 containing the anchor phrase ``mental health", compared to a random sample of tweets in 2020.} In the central 2D rank-rank histogram panel,
    phrases appearing on the right have higher rank in the mental health subset than in random tweets, while phrases on the left appeared more frequently in the random sample. The table to the right shows the words that contribute most to the divergence. For example, the phrase ``take care of" was the 112th most common 3-gram in random tweets posted during 2020, but it was the most common 3-gram in tweets containing ``mental health''. Note that when ranking 3-grams from mental health tweets, ``* mental health" and ``mental health *" phrases were removed for clarity.
    The balance of the words in these two subsets is also noted in the bottom right corner of the histogram, showing the percentage of total counts, all words, and exclusive words in each set. 
    See Dodds~\etal~\cite{dodds2020allotaxonometry} for a detailed description of our allotaxonometric instrument.}
    \label{fig:rankdiv_2020}
\end{figure*}

\begin{figure*}[!tp]
    \centering
    \includegraphics[width=\textwidth]{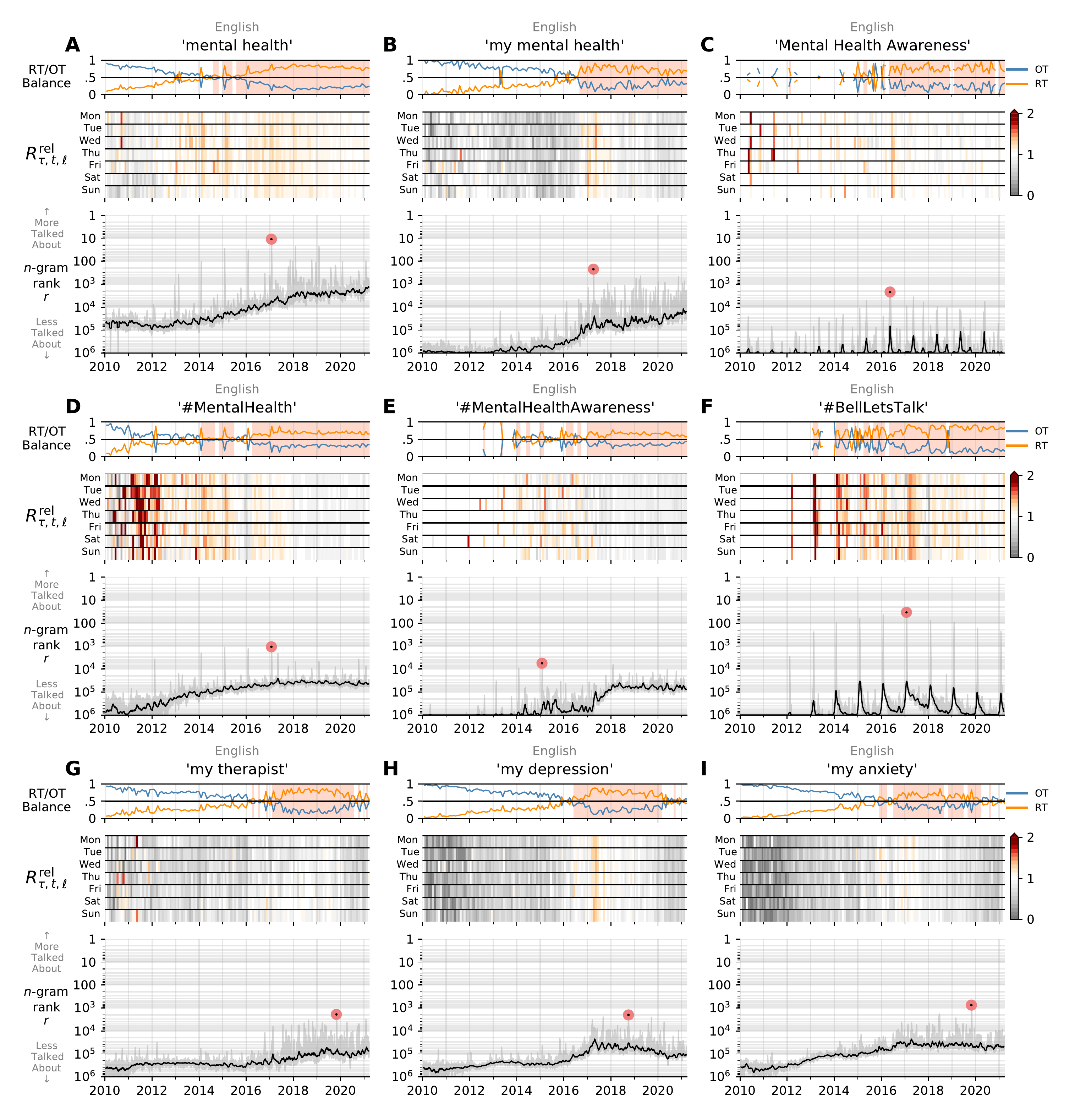} 
    \caption{\textbf{Contagiograms for mental health related $n$-grams.}  
    Phrases and hashtags related to the topic of mental health have grown in volume throughout the time period studied, as reflected by their popularity relative to all tweets.
    In each subplot, the top panel displays the monthly relative usage of each $n$-gram, indicating whether they appear organically in new tweets (OT, blue), or in shared retweets (RT, orange). The shaded area highlights time frames when the number of retweeted messages is higher than that of organic messages, suggesting social amplification~\cite{alshaabi2020storywrangler}. 
    The middle panel of each subplot shows the retweet usage of each $n$-gram relative to the background rate of retweets among all English tweets, with a heatmap for each day of the week. For these heatmaps, the color map is shown to the right, with darker red representing a higher relative rate of retweeting among these messages compared to general messages, and grey representing a higher rate of original messages. 
    The bottom panel shows the basic $n$-gram rank timeseries, with a month-scale smoothing of the daily values shown in black, and background shading in grey between the minimum and maximum rank of each week. 
    Note that phrase counts only reflect tweets that have been identified as messages written in English as discussed by Alshaabi \etal~\cite{alshaabi2020growing}.}
    \label{fig:MHcontagiograms}
\end{figure*}

\subsection{Narrative and Social Amplifications}
\label{subsec:narrative}

The increasing appearance of the phrase ``mental health” could be due to several factors. 
We analyze the corpus associated with the topic of ``mental health" using the $n$-grams and their relative frequency and rank values for each day, and compare the word usage in this subset to a random sample of messages on Twitter.

To compare differences in language usage, we use rank-turbulence divergence~\cite{dodds2020allotaxonometry}. With this method, we can examine the shift in language between the two samples of tweets. 
We aggregate $n$-gram counts for phrases found in tweets containing ``mental health'' over the span of each year, getting annual counts for each of these phrases. 

We do the same aggregation for a smaller random subset of Twitter data, aggregating yearly data for a one percent sample of the Decahose API. 
Fig.~\ref{fig:rankdiv_2020} highlights the results of rank divergence comparing the two subsets of messages across the year 2020. 

Each square histogram bin reflects the relative ranks for 3-word phrases in each respective subset. 
Bins to the right side contain 3-grams with relatively higher rank in the right subset than the left. 
The bins down the middle of the plot contain words with a similar rank in both subsets. 
The bands of bins on the bottom edges of these plots represent words that are exclusive to their respective side's dataset. 

The color of each bin correlates with the density of words contained in it, and the words appearing on the plot are randomly selected representatives from the bins on the outer edges. 
The table on the right shows the words that contribute most to the divergence of the two datasets, with small triangles indicating when a word is exclusive to one system.

When comparing $n$-grams from these subsets in Fig.~\ref{fig:rankdiv_2020}, we see that the mental health dataset, shown on the right side of the figure, includes language related to taking care of your physical and mental health, suicide prevention, men's mental health, social media, and personal time. 
These topics seem to have become more prominent in the year 2020 with people being at home and isolated during the COVID-19 pandemic, and with more awareness being brought to the relationship between social media and mental health. 
While we would expect to see pandemic-related phrases show up in 2020, these topics were equally mentioned across both samples, so they do not appear on either side of this histogram. 

Studies this year have shown that at the onset of the pandemic, Google searches for terms related to mental health increased initially, followed by a ``flattening out'' after stay-at-home orders were announced~\cite{jacobson2020flattening}.
It has also been recorded that in the time between March and July 2020, average phone screen time doubled to 5 hours per day and rates of depression increased by 90 percent~\cite{giuntella2021lifestyle}. 
While these figures cannot tell us everything about how language differs between subsets of conversation, they do provide a sense of the mental health topics individuals discussed in 2020.

To better understand the dynamics of phrases related to mental health, we explore ways in which these messages are spreading across Twitter. 
Tweets can be either posted as original content in a new message, or a user can retweet a message that another user has posted.

Organic messages show that users are writing their own content related to a topic, while retweeted messages show that this topic is being shared and spread to other groups of users; both are important means of contributing to conversation. 
Both organic messages (OT) and retweeted messages (RT) appear in our dataset and are included in the previous analyses, so it is important to also examine the proportion of messages that fall into these two categories. 

Fig.~\ref{fig:MHcontagiograms} shows ``contagiogram'' plots, as implemented by Alshaabi~\etal~\cite{alshaabi2020storywrangler}, which highlight the relationship between retweeted and organic content for a given $n$-gram on Twitter. 
The top panel of these plots shows the monthly relative usage of the specified $n$-gram, highlighting usage of organic messages in blue and shared retweets in orange. 
A shaded area in this top panel represents time periods when the number of retweeted messages surpasses that of organic messages, highlighting social amplification. 

The middle panel shows retweet usage of an $n$-gram, relative to the rate of all retweeting behavior across English Twitter, using a heatmap for each day of the week across the timeseries. 
In this heatmap, darker red shades represent a higher relative rate of retweets for the given $n$-gram compared to a random English $n$-gram on Twitter, and grey shades represent a higher rate of original messages. 
The bottom panel provides the rank timeseries of the $n$-gram, with a month-scale smoothing of the daily values shown in black. 
In Fig.~\ref{fig:MHcontagiograms}, we look at these contagiogram plots for a collection of key $n$-grams related to the discussion of mental health on Twitter.

Looking at English Twitter overall, the balance of messages was primarily organic until around 2017, when the practice of retweeting messages tipped the balance~\cite{alshaabi2020growing}. 
Around this same time, retweeted messages reach higher numbers than organic messages for most mental health related $n$-grams, as seen in the top panels of these subplots. 

Examining the heat map panels of these subplots, we observe a larger social amplification effect in hashtags related to mental health, highlighted by the darker red shades across the heatmaps. 
In recent years, however, these hashtags shift to more organic messages, with the heatmaps becoming more grey after around 2018.
The hashtag ``\#BellLetsTalk'' sees the most retweeted behavior of these hashtags, as well as an annual spike on the day of the event, followed by a substantial tail of conversation following this date.
On Mental Health Awareness Day (October 10th) of 2018, organic tweets referencing \#BellLetsTalk spiked, leading to the inversion of RT/OT in late 2018 that we see in Fig.~\ref{fig:MHcontagiograms}F.
We also see more original content containing self-disclosure phrases, such as ``my therapist" or ``my depression", as seen in the third row of $n$-grams which appear to have largely grey shades across the heatmaps.
These relationships suggest that users are sharing hashtags in order to spread awareness, and feel comfortable retweeting hashtags posted by others.
The public disclosure of private personal anecdotes, which helps to normalize conversation about personal struggles with mental health, is treated differently. 

Overall, our results suggest that a larger number of individuals feel comfortable making mental health disclosures publically, but they are amplified relatively less often than other types of mental health messages. 
Across the subplots, we see a substantial increase in the rank of all phrases/hashtags over time, with annual awareness days resulting in spikes corresponding to their given date each year. 
These findings offer evidence that understanding mental health conversations have increased substantially over time, reducing the stigma surrounding mental illness.

\section{Concluding Remarks}
\label{sec:conclusion}
In this project, we explored the conversation around mental health and its appearance on the social media platform Twitter. 
Using a collection of phrases, we examined how often the topic of mental health is discussed in tweets, finding that the 2-gram ``mental health” has increased in rank by nearly two orders of magnitude since 2012. 
We calculate the associated ambient happiness for the same time series, finding that happiness is largely effected by key dates and has generally decreased over the past decade. 

Compiling a new dataset of $n$-grams found in the subset of tweets mentioning ``mental health”, we analyzed text associated with this specific term, finding the top $n$-grams related to the topic and their usage rates. 
We examine the the language in this conversation across years, finding topics that emerged over the past year since the pandemic began. 

Comparing usage rates of retweeted content and original content, we find that common ``awareness” messages are being amplified on the social media platform, while personal self-disclosing statements are being seen more in organic, originally authored content. 
These results provide valuable insight into how the discussion of mental health has changed over time, and suggests that more awareness and acceptance has been brought to the topic compared to past years. 

We acknowledge that using Twitter as a data source for this research has many limitations, as its user base is not a broadly representative sample of the human population. A study by the Pew Research Center~\cite{perrin2019share} shows that as of June 2019, a only 22 percent of all US adults reported using Twitter, smaller for example than the 69 percent who use Facebook. 
The age breakdown of users is also skewed, with 38 percent of 18-29 year-olds using Twitter while only 17 percent of 50-64 year-olds use the site. 

While demographics of race are fairly uniform (21 percent of white adults, 24 percent of black adults, and 25 percent of Hispanic adults), the platform is more often used by individuals with a college degree (32 percent) living in an urban area (26 percent)~\cite{perrin2019share}. 
We also recognize that a portion of Twitter accounts are run by businesses, institutions, and other organized groups, rather than simply individual people. 
These corporate accounts, such as ``@Bell\_LetsTalk'', would have more of a pattern and agenda to their posted tweets, and there is not currently a way to filter out these messages. 
Due to these complexities of the Twitter user base, care must be taken when interpreting findings based on tweets.

These limitations could be addressed in future studies by expanding the data sources, e.g., by looking to other available online sites such as Reddit, Instagram, or Facebook, whose user bases differ in some regards. 
Turning away from social media, one could examine clinical records for cases of diagnosed mental illness, 
analyzing the language and positivity of physician notes.
Rather than looking at simply the messages of this social media platform, this work could be expanded to address the conversation on a network scale, determining how interactions between users impact the discourse. 

The work presented here is also limited to the anchor phrase ``mental health", and thus could be leaving out conversation related to the topic. To further enrich these findings, future work could expand the existing mental health dataset to include tweets with additional anchor $n$-grams, although a method for determining these anchors would be necessary.  

We believe the results presented here provide useful texture regarding the growing conversation around mental health on Twitter, and evidence that more people are contributing to this conversation on the public social media platform than ever before.
Public health campaigns aiming to reduce stigma surrounding mental health can leverage success stories to improve their messaging.
As this conversation continues to grow, and perhaps becomes more normalized, it will be useful to examine the language or events that could be contributing to these shifts.

\acknowledgments 
The authors are grateful for the computing resources provided by the Vermont Advanced Computing Core and financial support from the Massachusetts Mutual Life Insurance Company. We thank many of our colleagues at the Computational Story Lab for their feedback on this project.  

\bibliography{references}

\clearpage

\newwrite\tempfile
\immediate\openout\tempfile=startsupp.txt
\immediate\write\tempfile{\thepage}
\immediate\closeout\tempfile

\setcounter{page}{1}
\renewcommand{\thepage}{S\arabic{page}}
\renewcommand{\thefigure}{S\arabic{figure}}
\renewcommand{\thetable}{S\arabic{table}}
\setcounter{figure}{0}
\setcounter{table}{0}




\end{document}